\documentclass[a4paper,fleqn]{cas-sc}

\usepackage[]{natbib}

\usepackage{bm}

\def\tsc#1{\csdef{#1}{\textsc{\lowercase{#1}}\xspace}}
\tsc{WGM}
\tsc{QE}

\begin{document}
\let\WriteBookmarks\relax
\def\floatpagepagefraction{1}
\def\textpagefraction{.001}

\shorttitle{Theoretical approaches to liquid Helium}    

\shortauthors{J. Boronat}  

\title [mode = title]{Theoretical Approaches to Liquid Helium}  



%

\author[]{Jordi Boronat}[orcid=0000-0002-0273-3457
]






\affiliation[]{organization={Departament de F\'\i sica, Universitat 
Polit\`ecnica de Catalunya},
            addressline={Campus Nord B4-B5}, 
            city={Barcelona},
            postcode={08034}, 
            country={Spain}}










\begin{abstract}
In this article, we review the main theoretical methods applied to the study of 
liquid Helium adopting a microscopic approach, that is, starting from the 
many-particle Hamiltonian of the system. Following an introduction on the first 
early approaches, we discuss two main issues. In the first one, we report the 
main ingredients of a theoretical approach based on the variational method, 
with the discussion of the high accuracy obtained with them and the related 
progress in the design of highly accurate trial wave functions. In the second 
part, the main stochastic methods used in this study are briefly discussed. 
Altogether reflects the strong links between the study of liquid Helium and the 
progress in the development of new and extremely powerful approaches to deal 
with one of the most strongly quantum many-body systems in Nature.
\end{abstract}




\maketitle

\underline{Key points/Objectives}
\begin{itemize}
\item Main physical properties of liquid Helium, both in its superfluid and 
normal states.

\item Early phenomenological theoretical approaches to account for the main 
experimental data.

\item Development of theoretical approaches from a microscopic view: 
variational and correlated basis function theories.

\item Theoretical study based on stochastic methods: quantum Monte Carlo.

\end{itemize}

\section{Introduction}\label{introduction}

Liquid Helium was first liquefied by Kammerlingh Onnes in 
1908~\citep{Onnes1908} and, since then, it has been one of the most studied 
quantum systems. Two of its isotopes are stable, $^4$He and $^3$He. In 
naturally produced Helium, the fraction of $^3$He is extremely small, 1 part in 
10$^6$, and is produced from radioactive decay of tritium. Both isotopes remain 
in liquid state even in the limit of zero temperature, manifesting in this way 
their intrinsic quantum nature. Contrarily to classical theory, that predicts 
that any material becomes solid in that limit, quantum mechanics explain this 
fail of the classical approach by considering the zero-point kinetic energy. 
The combination of the light mass of Helium atoms and their weak attraction 
produces the unique opportunity of observing the most paradigmatic quantum 
liquid. $^4$He and $^3$He have different masses but the most crucial difference 
between both isotopes affects their behavior as quantum many-body systems: 
$^4$He is composed by an even number of spin $1/2$ particles and thus, it is a 
boson, whereas $^3$He has an odd number of constituents and behaves as a 
fermion. As we will discuss in this article, the different quantum statistics 
of the two Helium isotopes produces dramatic differences between them.

In $^4$He, the specific heat shows a large excess at a temperature of 
$2.17$ K, signaling the emergence of a phase transition, first observed by 
Keesom and Clusius~\citep{Keesom1932}. The specific heat shows a shape similar 
to 
the $\lambda$ Greek letter and is now universally known as the $\lambda$ phase 
transition, with a critical temperature $T_c=2.17$ K. This second-order 
transition separates two liquids with different properties and were termed 
liquid He I and liquid He II for $T>T_c$ and $T<T_c$, respectively. In 1937, P. 
Kapitza~\citep{Kapitza38}, from one side, and J. F. Allen and A. D. 
Misener~\citep{Allen38}, on the other, published back to back papers announcing 
the discovery of superfluid $^4$He in the Helium II phase. For an insightful 
historical analysis of the role played by the two independent teams on that 
discovery, we recommend the paper by Balibar~\citep{Balibar07}. Below 
$T_\lambda$, $^4$He flows with vanishingly small viscosity producing a set of 
surprising effects such as the fountain effect, absence of boiling due to the 
large thermal conductivity, second sound, etc.~\citep{Wilks67}.

\section{Early approaches}\label{Early}

In the early theoretical approaches to understand the extraordinary behavior of 
liquid Helium around the $\lambda$ point, four names appear above many others: 
Laszlo Tisza, Lev Landau, Fritz London and Nikolai Bogoliubov. Tisza and Landau 
worked in a quantum hydrodynamic approach with the main  idea of Helium as a 
liquid composed by two independent liquids (velocity fields). This is the basis 
of the two-fluid model first introduced by Tisza~\citep{Tisza38} and then 
completed by Landau~\citep{Landau41}. The two-fluid model assumes that the total 
density is the sum of two partial densities: the normal and the superfluid. 
Below $T_\lambda$, the superfluid component dominates, thus explaining the 
superfluid properties of Helium II. Increasing the temperature, at $T \ge 
T_\lambda$, the superfluid density turns to be zero and the total density equals 
the normal one.   

Landau~\citep{Landau41} introduced the concept of quasiparticle to postulate the 
shape of the excitation spectrum. Instead of single-particle excitations, the 
system presents collective modes that at low momenta are termed phonons, 
in resemblance with the excitation spectrum of a crystal. At low momenta, the 
excitations are linear with the momenta, the slope being the speed of sound. At 
larger momenta, Landau defined another type of collective excitations which were 
named rotons, since where initially related to the presence of quantized 
vortices. Landau predicted that below a certain velocity, termed critical 
velocity, the system cannot be excited and thus liquid Helium flows without 
friction. Combining phonons and rotons, Landau~\citep{Landau47} was able to draw 
the excitation spectrum of Helium before any measure of it was made. 

The hydrodynamic theory by Tisza and Landau did not take into account the 
quantum statistics of $^4$He because they believed that this feature was not 
relevant to understand the superfluidity and the $\lambda$ phase transition. On 
the contrary, London~\citep{London38} argued that a Bose gas experiences a 
second-order phase transition that can explain, at least qualitatively, the 
$\lambda$ transition in $^4$He. Applying the critical temperature derived in 
the Bose-Einstein statistics,
\begin{equation}
k_B T_{\text{BEC}} = \frac{2 \pi \hbar^2}{m} \, \left( \frac{n}{g_{3/2}(1)} 
\right)^{2/3} \ ,
\label{tbec}
\end{equation}
with $g_{3/2}(1)\simeq 2.612$ and $n$ the density, to the case of Helium London 
obtained $3.1$ K, not so far from $T_\lambda$. Landau always ignored London 
approach to the problem~\citep{Balibar07} because with his hydrodynamic theory he 
was able to account for many experimental facts. However, we know now that the 
quantum statistics of $^4$He is crucial to understand superfluid $^4$He. When 
samples of $^3$He were produced, we learned that  $^3$He, 
which is a fermion, do not show the $\lambda$ phase transition. 

Bogoliubov, in 1947~\citep{Bogoliubov47}, was the first to connect the elementary 
excitation spectrum with a Bose-Einstein gas with repulsive but small 
interatomic interaction. He proved that the lowest-energy excitations are 
collective modes (phonons) that at low momenta are proportional to the speed of 
sound, in agreement with Landau theory. The Bogoliubov excitation 
spectrum~\citep{Stringari_book}
\begin{equation}
\varepsilon(p) = \left[ \frac{g n}{m} p^2 + \left( \frac{p^2}{2 m} \right)^2 
\right]^{1/2}   \ ,
\label{bogoli}
\end{equation}
with $g$ the interaction strength, predicts a critical velocity which is always 
equal to the speed of sound. Bogoliubov theory was again qualitatively correct 
but Helium is not a rarefied gas but a liquid with strong interparticle 
interactions.

\section{Microscopic theories}\label{micro}

We discuss first liquid $^4$He, $^3$He was obtained later in time due to its 
complex experimental production. As we commented in the previous Section, 
Bogoliubov~\citep{Bogoliubov47} introduced field theoretical models in the 
theoretical description of liquid $^4$He, but his approach was only applicable 
to dilute Bose gases and this is far form the real properties of Helium. This 
is specially evident from the explicit assumption in Bogoliubov analysis that 
all the particles of the system are in the Bose-Einstein condensate, whereas in 
$^4$He only ~7\% of them are really in this state due to the unavoidable role 
played by atomic correlations. In subsequent work, Beliaev~\citep{Beliaev58} 
introduced Green functions $G_{\alpha,\beta}(q,\omega)$ in the formalism, deriving a general equation for 
them  at zero temperature. The quantum field theory 
for a Bose fluid was reanalyzed by Hugenholtz and Pines~\citep{Pines59} by the 
introduction of the chemical potential $\mu$ to guarantee the number 
conservation of the theory. Importantly, they removed the constraint of a full 
condensate, approaching better the characteristics of liquid Helium. An 
important result of this approach was the calculation of the energy of a Bose 
gas incorporating the first corrections to the Hartree-Fock energy
\begin{equation}
\frac{E}{N} = 4\pi na^3 \left[ 1 + \frac{128}{15\sqrt{\pi}}\sqrt{na^3} 
+ \frac{8(4\pi-3\sqrt{3})}{3}na^3 \ln(na^3) + \ldots \right] \ ,
\label{enexp}
\end{equation}
the energy per particle $E/N$ being in units of $\hbar^2/(2 m a^2)$. Hugenholtz 
and Pines~\citep{Pines59} proved that, beyond the expansion terms in 
Eq.(\ref{enexp}), additional contributions will depend on the specific shape of 
the interatomic potential and not only on the $s$-wave scattering length $a$. 
The coefficient of the $(na^3)^{3/2}$ term was first calculated by Lee, Huang
and Yang \citep{Lee57}, while the coefficient of the last term was first 
obtained by Wu \citep{Wu59}. Both of them were originally derived for hard 
spheres, but it was shown that the same expansion is valid for any repulsive 
potential with scattering length $a$ \citep{Pines59}.  Equation (\ref{enexp}) 
has been very useful in the study of dilute Bose gases, composed by alkali 
atoms, due to their extreme diluteness~\citep{Stringari_book}. However, liquid 
Helium is a strongly interacting quantum many-body system in which any 
expansion in terms of the gas parameter $na^3$ is completely useless.

The first theoretical approach based on the wave function of the $N$-body 
quantum system was made by Feynman~\citep{Feynman53}. This work opened a new way 
to deal with the properties of Helium based on a microscopic approach. Starting 
from general arguments, that included the symmetry of the $^4$He atoms and 
their hard-core interactions at short distances, Feynman argued that the 
lowest-energy excitations have a collective behavior since single-particle ones 
have always higher energy. Based on these considerations, he wrote the wave 
function of the excited state of the $N$-body fluid as
\begin{equation}
\Psi(\bm{r}_1,\ldots,\bm{r}_N) = \prod_{i=1}^{N} f(\bm{r}_i) \ 
\Phi(\bm{r}_1,\ldots,\bm{r}_N) \ ,
\label{feynman1}
\end{equation}
with $\Phi(\bm{r}_1,\ldots,\bm{r}_N)$ the ground-state wave function. Equation 
(\ref{feynman1}) is a variational approach to the problem and thus the energy 
of the excitation is an upper bound to the exact energy. He obtained the lowest 
excitation energy
\begin{equation}
\hbar \omega_{\text F}(q)= \frac{\hbar^2 q^2/2 m}{S(q)} \ ,
\label{feynman2}
\end{equation}
with $S(\bm{q})=1/N \langle \rho^\dagger(\bm{q}) \rho(\bm{q}) \rangle$ the 
static structure factor, $\rho(\bm{q})=\sum_i \exp(-i \bm{q} \cdot \bm{r}_i)$ 
being the density fluctuation operator. When $\bm{q} \to 0$, $S(q)=\hbar q /(2 
m c)$, with $c$ the speed of sound. In this way, Feynman recovers the phonon 
relation previously found by Landau and Bogoliubov, $\hbar \omega = c q$, that 
is, a linear behavior.  At large momenta, the static structure factor tends to 
1 and so one recovers the excitation energy of a free particle $\hbar^2 
q^2/2m$, quadratic with $q$. At intermediate $q$ values, $S(q)$ shows a peak 
related to the mean interparticle distance and thus the excitation energy 
(\ref{feynman2}) shows a local minimum that we can identify with the roton. 
Qualitatively, the Feynman spectrum is correct up to momenta not much larger 
than the one of the roton but the roton energy is nearly two times larger than 
the experimental one. The minimum energy (\ref{feynman2}) is obtained with a 
function $f(r)=\exp(i \bm{q} \cdot  \bm{r})$, so the total wave function 
(\ref{feynman1}) is given by
\begin{equation}
\Psi(\bm{r}_1,\ldots,\bm{r}_N) = \prod_{i=1}^{N} e^{i \bm{q}\cdot \bm{r}_i} \ 
\Phi(\bm{r}_1,\ldots,\bm{r}_N) = \rho^\dagger(\bm{q})  
\Phi(\bm{r}_1,\ldots,\bm{r}_N) \ ,
\label{feynman3}
\end{equation}
that emerges as the creation of a collective mode of momentum $\bm{q}$ acting 
on the ground-state, that is not changed by the excitation.

Feynman and Cohen~\citep{Feynman56} improved the initial theory by introducing 
in the wave function two-body correlations. Their idea is that when one 
particle moves in the fluid its neighbors \textit{feel} the movement or, in 
other words, the movement of each particle depends on its local environment. 
They termed these correlations as backflow correlations. Explicitly, the wave 
function of the collective excitation turns to
\begin{equation}
\Psi(\bm{r}_1,\ldots,\bm{r}_N) = \prod_{i=1}^{N} \left[ e^{i \bm{q}\cdot 
\bm{r}_i}  \sum_{j \ne i} e^{i g(\bm{r}_i-\bm{r}_j)} \right]     
\Phi(\bm{r}_1,\ldots,\bm{r}_N)  \ .
\label{back1}
\end{equation}
The function $g(r)$ in Eq. (\ref{back1}) couples the movement of the excited 
particle $i$ with the neighboring ones. In Ref.~\citep{Feynman56}, it is taken 
as a dipolar term, $g(r)= \alpha \bm{q} \cdot \bm{r}/r^3$ and the imaginary 
exponential containing $g(r)$ is linearized to make the calculation of the 
energy simpler. With this correction, the difference between the Feynman roton 
energy (\ref{feynman2}) and the experimental one is reduced in a factor $\sim 
2$ but still there is a significant difference.

\begin{figure}
\begin{center}
\includegraphics[width=0.7\linewidth]{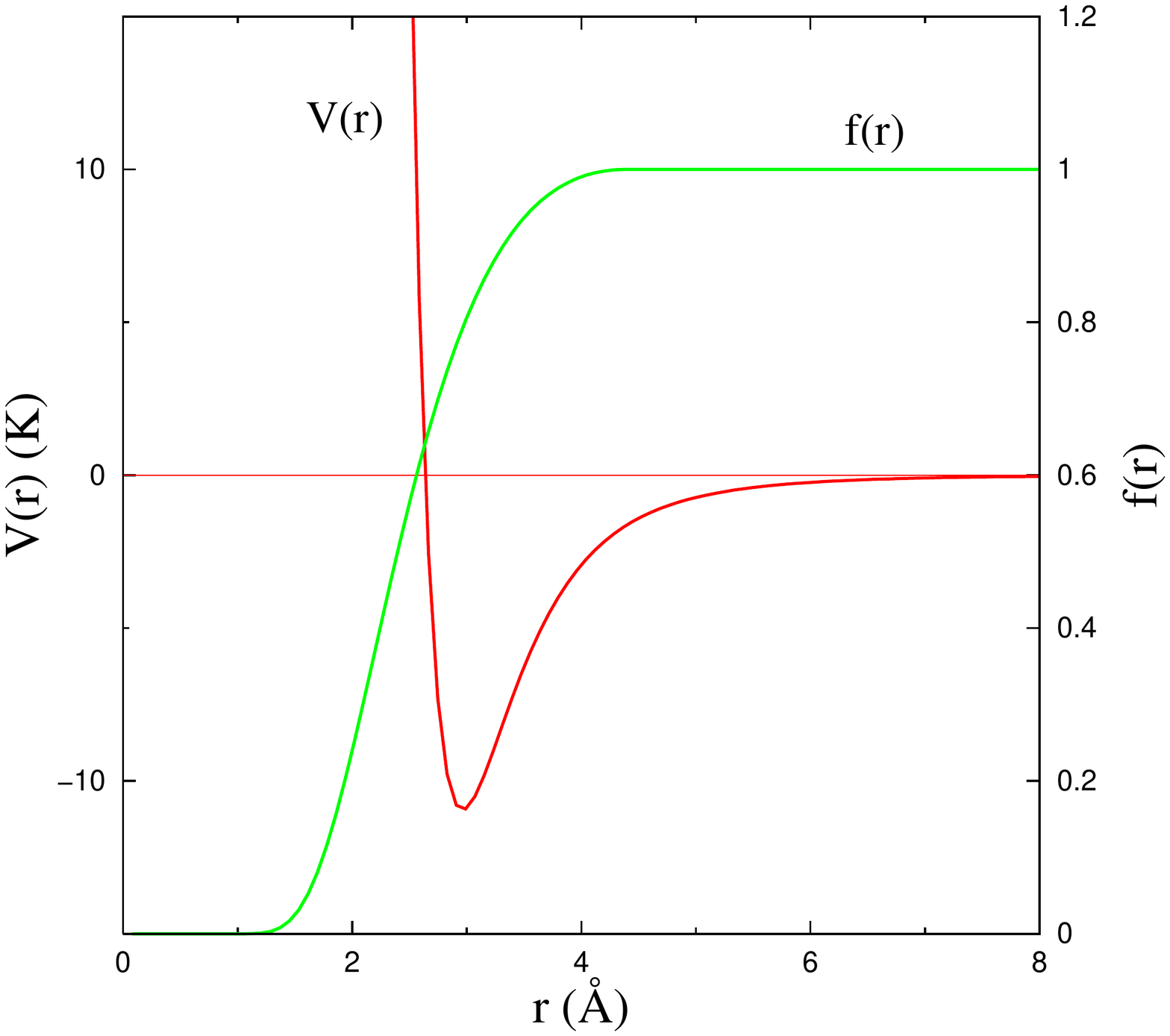}
\end{center}
\caption{He-He interatomic potential $V(r)$ and the two-body correlation factor 
$f(r)$.    }
\label{fig-fr}
\end{figure}

To reproduce as better as possible the elementary excitation spectrum of liquid 
$^4$He is not the only goal of a microscopic theory. It is fundamental to have 
an accurate description of the ground state of the system. The most fruitful 
approach to account for the equation of state of liquid $^4$He has been the 
variational theory since standard perturbative approaches are not applicable 
due to the hard-core of the He-He interaction at short distances. The 
Hamiltonian of the $N$-particle system is
\begin{equation}
H=-\frac{\hbar^2}{2m} \sum_{i=1}^{N} \bm{\nabla}_i^2 + \sum_{i<j}^{N} V(r_{ij}) 
\ , 
\label{hamiltonian} 
\end{equation}
with $V(r)$ the pair interatomic potential. One can add to 
Eq. (\ref{hamiltonian}) three-body potentials, whose analytic form is not well 
known. However, its contribution to the energy seems to be not relevant and 
the results can even worsen~\citep{Boro94}. In the description of the bulk phases, 
one always assumes the thermodynamic limit: $N  \to \infty$, $\Omega \to 
\infty$, and the density $\rho=N/\Omega$ is finite. For a trial wave function 
$\Psi$, not orthogonal to the ground state, the Ritz's variational principle of 
quantum mechanics states that
\begin{equation}
E=\frac{ \langle \Psi | H | \Psi \rangle}{\langle \Psi|\Psi \rangle} \ge E_0 \ ,
\label{ritz}
\end{equation}
with $E_0$ the ground-state energy of the system. The wave function $\Psi$ must 
be symmetric under the interchange of particles and approaches zero when 
$r_{ij} \to 0$. Also, when a set of particles is significantly far way from 
the rest of the system the total wave function factorizes, i.e., the cluster 
property is fulfilled. The first model wave function that comes to mind is a 
symmetrized product of single-particle wave functions which, for a bulk phase, 
are plane waves. However, this model clearly violates both crucial properties.  
Bijl~\citep{Bijl} and Jastrow~\citep{Jastrow} proposed a model to account with 
the restrictions imposed by a strongly correlated fluid,
\begin{equation}
\Psi(\bm{r}_1,\ldots,\bm{r}_N)= \prod_{i=1}^N g(r_i) \, \prod_{i<j}^N f(r_{ij}) 
\ ,
\label{jastrow}
\end{equation}
with $f(r)$ a two-body correlation factor and $g(r)$ a one-body function that, 
in the bulk phase, is simply 1 (ground-state). The function $f(r)$ plays an 
essential role in the study of Helium and satisfies two main properties: 
it becomes zero, approaching the hard-core of the potential, and one at large 
distance, according to the cluster property. In Fig.\ref{fig-fr}, we show the 
characteristic shape of the potential and the two-body correlation factor. 

The energy per particle for a Bijl-Jastrow wave function is readily obtained as
\begin{equation}
\frac{E}{N}= \frac{1}{2} \rho \int d \bm{r} \left[ V(r)- \frac{\hbar^2}{2m} 
\bm{\nabla}^2 \ln f(r) \right]  \ ,
\label{energyjf}
\end{equation}
with $g(r)$ the two-body radial distribution function
\begin{equation}
g(r_{12}) = \frac{N(N-1)}{\rho^2} \frac{\int d\bm{r}_3 \ldots d\bm{r}_N 
|\Psi(\bm{r}_1,\ldots,\bm{r}_N)|^2}{\int d\bm{r}_1 \ldots d\bm{r}_N 
|\Psi(\bm{r}_1,\ldots,\bm{r}_N)|^2} \ ,
\label{gr}
\end{equation}
which is proportional to the probability of finding two particles at distance 
$\bm{r}$. Accessing to the function $g(r)$ allows for the calculation of the 
energy of the system and further properties. However, as Eq. (\ref{gr}) shows, 
its estimation is very difficult due to the multidimensional integrals that 
need to be carried out. A fruitful approach in the study of liquid Helium has 
been the hyppernetted chain theory (HNC) based on the summation of the series 
in terms of the function $h(r)=f^2(r)-1$ in which $g(r)$ can be expressed. This 
function $h(r)$ is different from zero only at short distances and allows for a 
cluster expansion. $g(r)$ is given exactly by
\begin{equation}
g(r)= f^2(r) \, \exp (N(r)+E(r)) \ ,
\label{nodals}
\end{equation}
where $N(r)$ and $E(r)$ are functions summing up specific terms of the series, 
known as nodal and elementary diagrams, respectively. Nodal diagrams can be 
calculated by solving the HNC integral equation
\begin{equation}
N(r_{12})=\rho \int d \bm{r}_3 [ g(r_{13})-1 ] [ g(r_{32})-N(r_{32}) -1 ]  \ .
\label{hnc}
\end{equation}
On the contrary, elementary diagrams cannot be summed up and, thus, only 
approximated in some way~\citep{Usmani82,Fabrocini82,Clements93}. Therefore,  
Eq. (\ref{nodals}) cannot be evaluated in an exact and closed form using this 
theory.   

The variational upper bound to the exact energy depends on the trial wave 
function. One can guess an analytic model for it, with a set of variational 
parameters to be optimized. Otherwise, one can approach the optimization via a 
functional derivative, normally written in terms of the two-body distribution 
function $g(r)$,
\begin{equation}
\frac{\delta E[g(r)]}{\delta g(r)} = 0 \ ,
\label{optim1}
\end{equation}
with $E[g(r)]$ the energy of the system. This can be written only in terms of 
$g(r)$ by using Eq. (\ref{nodals}),
\begin{equation}
E[g(r)] = \frac{\rho}{2} \int d \bm{r} g(r) V(r) + \frac{\hbar^2 \rho}{2m} \int 
d \bm{r} \left[ \left(\bm{\nabla} \sqrt{g(r)} \right)^2 + \frac{1}{4} g(r) \bm{\nabla}^2 
E(r) \right] -\frac{\hbar^2}{8m}\frac{1}{2 \pi^3 \rho} \int d \bm{k} k^2 
\frac{(S(k)-1)^3}{S(k)}    \ .
\label{optim2}
\end{equation}
Imposing the minimization condition (\ref{optim1}), one arrives to the 
Euler-Lagrange differential equation for the optimal 
correlation~\citep{Lantto77,Kallio77}
\begin{equation}
-\frac{\hbar^2}{m} \bm{\nabla}^2 \sqrt{g(r)} + ( V(r) + W(r) ) \sqrt{g(r)} = 0 
\ .
\label{eulerlagrange}
\end{equation}
This equation needs to be solved iteratively~\citep{Jackson79} because the 
potential 
$W(r)$ depends on the correlation function. Neglecting the contribution of the 
elementary diagrams, this induced potential is better written in 
reciprocal space as 
\begin{equation}
\tilde{W}(q) = -\frac{\hbar^2 q^2}{4m} \frac{(S(q)-1)(2S(q)+1)}{S(q)^2} \ , 
\end{equation}
with $S(q)$ the static structure factor,
\begin{equation}
S(q)= \frac{1}{N} \langle \rho_{\bm{q}} \rho_{-\bm{q}} \rangle = 1 + \rho \int 
d \bm{r} e^{i \bm{q}\cdot\bm{r}} (g(r)-1) \ .
\label{sk}
\end{equation} 
At zero temperature, the function $S(q)$ satisfies $S(0)=0$ and shows the 
phononic behavior at small $q$
\begin{equation}
\lim_{q \to 0} S(q)=\frac{\hbar q}{2 m c} \ ,
\label{sqpetit}
\end{equation}
with $c$ the speed of sound. This behavior at small $q$ produces a long-range 
behavior of the two-body correlation factor~\citep{Feenberg}
\begin{equation}
\lim_{r \to \infty} f(r)= 1-\frac{m c }{2 \pi^2 \hbar \rho r^2} \ .
\label{frlong}
\end{equation}
Remarkably, the optimal $f(r)$ obtained from the Euler-Lagrange equation 
(\ref{eulerlagrange}) satisfies the phonon behavior 
(\ref{frlong})~\citep{Castillejo79}, even if the speed of sound that derives 
form that differs from the experimental one.

At zero temperature, the superfluid fraction of $^4$He is one but the 
occupation of the zero-momentum state, that is the Bose-Einstein condensate, is 
significantly smaller than one due to the relevance of correlations. 
Microscopically, the condensate fraction $n_0$, and the momentum distribution 
in general, can be calculated through the one-body density matrix
\begin{equation}
\rho(r_{1 1^\prime})= \Omega \, \frac{\int d\bm{r}_2 \ldots 
d\bm{r}_N 
\Psi(\bm{r}_1,\ldots,\bm{r}_N)   \Psi(\bm{r}_{1^\prime},\ldots,\bm{r}_N)  
}{\int d\bm{r}_1 \ldots d\bm{r}_N 
|\Psi(\bm{r}_1,\ldots,\bm{r}_N)|^2} \ ,
\label{onebody}
\end{equation}
normalized as $\rho(0)=1$. Fourier transforming $\rho(r)$ one gets the momentum 
distribution
\begin{equation}
n(q) =  n_0 \delta(q) + \rho \int d \bm{r} (\rho(r)-n_0) e^{i \bm{q} \cdot 
\bm{r}} \ ,
\label{nk}
\end{equation}
with $n_0 = \lim_{r \to \infty} \rho(r)$ the condensate fraction. The momentum 
distribution (\ref{nk}) can be measured using deep inelastic neutron scattering 
at large momentum transfer. The condensate fraction is quite small $n_0 \sim 
0.07$ and its estimation, from the dynamic structure factor $S(\bm{q},\omega)$, 
 requires from an accurate analysis of final state effects~\citep{Glydebook}.
A variational estimation of the one-body density matrix and the condensate 
fraction can be made using a  HNC cluster expansion similar to the one 
developed for $g(r)$~\citep{Fantoni78}. 

With the best Jastrow factor, the upper bound to the energy provided by the variational 
calculation is still $~15$\% above the exact quantum Monte Carlo results (see 
next Section) for the same interatomic potential and at densities close to the 
equilibrium point. An important advance on the improvement of the variational 
energy was the introduction of three-body correlations in the trial wave 
function,
\begin{equation}
\Psi_{\text{JT}}(\bm{r}_1,\ldots,\bm{r}_N)= \prod_{i<j}^N f(r_{ij}) \, 
\prod_{i<j<k}^N f_3(r_{ij},r_{ik},r_{jk}) 
\ ,
\label{triplet}
\end{equation}
which corrects  the description of correlation effects at short and medium 
interparticle distances. The relevance of these correlations in the study of 
liquid Helium derives from its relatively large density, which drives 
this liquid into the realm of one of the most correlated quantum 
systems in Nature. The origin of the function $f_3$, and its specific 
functional form, can be derived in different ways: introducing a momentum 
dependence in the two-body correlation factor~\citep{Pandha78} or by searching 
the first correction to the Jastrow factor by linearizing the imaginary-time 
Schr\"odinger equation~\citep{borobook}.  It adopts the form
\begin{equation}
f_3(r_{ij},r_{ik},r_{jk})= \exp \left(-\frac{1}{2} \, q(r_{ij},r_{ik},r_{jk}) \right) \ ,
\end{equation}
where
\begin{equation}
q(r_{ij},r_{ik},r_{jk})=\sum_l \sum_{\text{cyc}} \xi_l(r_{ij}) \xi_l(r_{ik}) 
P_l(\hat{r_{ij}} \cdot \hat{r_{ik}}) \ ,
\label{f3form}
\end{equation}
with $l$ the angular momentum and $P_l$ the Legendre polynomial. The most 
important, and most used contribution, corresponds to $l=1$. The pair function 
$\xi(r)$ in Eq. (\ref{f3form}) is assumed to be a short-range correlation factor
\begin{equation}
\xi(r) = \sqrt{\lambda} \, \exp \left[ - \left( \frac{r-r_t}{r_\omega} 
\right)^2 \right] \ ,
\label{xi}
\end{equation}
with $\lambda$, $r_t$, and $r_\omega$ variational parameters to be optimized. 
The HNC equations can be generalized to allow for the inclusion of three-body correlations in the 
diagrammatic expansions~\citep{Usmani82b}. The 
energies obtained so far reduces in a $\sim 70$\% the difference between the 
Jastrow upper bound and the exact values, pointing to its importance for a 
right microscopic description of liquid $^4$He.

The theoretical description of liquid Helium is not complete without understanding the dynamics of the system.
The dynamic structure function $S(\bm{q},\hbar \omega)$ contains the maximum information about it and is experimentally accessible by inelastic scattering through
\begin{equation}
\frac{\partial^2 \sigma}{\partial \Omega \, \partial \hbar\omega}=b^2 \left( \frac{q_1}{q_0} \right) S(\bm{q},\hbar \omega)
\ ,
\label{crosssection}
\end{equation}
with the left part being the double differential cross section. In Eq. (\ref{crosssection}), $b$ is the scattering length of the nucleus, and $\bm{q}_0$ and $\bm{q}_1$ are the initial and final momentum of the neutron. The momentum and energy transferred to the liquid are $\bm{q}=\bm{q}_1-\bm{q}_0$ and $\hbar \omega$, respectively.
At zero temperature, and for a $N$ particle system,
\begin{equation}
S(\bm{q},\hbar \omega)=\frac{1}{N} \sum_n |\langle \Psi_n | \rho(\bm{q}) | \Psi_0 \rangle |^2 \delta(\hbar \omega -(E_n-E_0)) \ ,
\label{sqw}
\end{equation}
where the sum is extended to all the excited states and $\rho(\bm{q})$ is the density fluctuation operator. To calculate the dynamic response one invokes linear response theory, where the change in the density under the action of an external tiny time-dependent potential is
\begin{equation} 
\delta \rho(\bm{r},t)= \int d\bm{r}' dt' \chi(\bm{r},\bm{r}',t-t') \, \delta V_{\text{ext}}(\bm{r}',t') \ ,
\label{linres}
\end{equation}
with $\chi$ the density-density response function. The dynamic response can be obtained from the imaginary part of the Fourier transform of $\chi$,
\begin{equation}
S(\bm{q},\hbar \omega)=-\frac{1}{N \pi} \Im \left[ \int d\bm{r} d\bm{r}' e^{i 
\bm{q}\cdot(\bm{r}-\bm{r}')} \chi(\bm{r},\bm{r}',\hbar \omega) \right] \ ,
\label{linres2}
\end{equation}
and the evolution in time of $\delta \rho(\bm{r},t)$ from the least-action 
principle $\delta S=0$. The calculation of the dynamic response of liquid 
Helium has been the object of intense work for many years. Regular perturbation 
theory does not work because of the hard-core of the He-He interaction and then 
it is necessary to work out the theory  using correlated basis functions (CBF). 
From pioneering work by Feenberg and Jackson~\citep{Jackson62} and Lee and 
Lee~\citep{Lee75}, the major progress has been achieved by dynamic many-body 
theory developed by Krotscheck and Campbell~\citep{Campbell09,Campbell15}. 
Recently, a comparison between very accurate experimental data, on the dynamic 
response and the excitation spectrum, and theoretical results obtained with 
dynamic many-body theory has shown an impressive agreement~\citep{Beauvois18}.

 Until here, we have focused our discussion on the studies of bosonic liquid 
$^4$He. But, Helium offers another stable isotope, $^3$He, which is a fermion. 
At very low temperatures, it remains in a liquid state that becomes superfluid 
at $\sim 2$ mK. In the following, we refer to normal non-superfluid $^3$He 
since superfluid $^3$He is a very different subject which would require a 
completely different scope. The variational wave function incorporates the 
Fermi statistics in the well-known Jastrow-Slater form
\begin{equation}
\Psi(\bm{r}_1,\ldots,\bm{r}_N)=  \prod_{i<j}^N f(r_{ij}) 
\, \Phi(\bm{r}_1,\ldots,\bm{r}_N)
\ ,
\label{jasslater}
\end{equation}
with $\Phi$ an antisymmetric wave function, usually taken as the product of 
Slater determinants $\Phi=D_\uparrow D_\downarrow$ since the interaction is 
spin independent. Variational theory, based on cluster expansion and summation 
of proper diagrams, to calculate the pair distribution function $g(r)$ is now 
more involved than in Bose fluids. The extension of HNC theory is known as 
Fermi-hypperneted chain (FHNC) and the convergence of the series summation was 
proved in the seventies of the past century~\citep{Fantoni75,Krotscheck75}. This 
theory was reviewed in depth by Rosati~\citep{Rosati81} and a modern update was 
carried out by Krotscheck~\citep{Krotscheck00}. As in the case of $^4$He, the 
inclusion in the trial wave function (\ref{jasslater}) of triplet correlations 
reduces significantly the difference with quantum Monte Carlo estimations. Even 
with the introduction of three body factors into the variational model, the 
Jastrow-Slater model uses Slater determinants with free single-particle 
orbitals. A relevant improvement of the variational model is achieved by 
modifying these orbitals to include, in an approximate way, the effect of 
correlations on them. This is usually made with backflow correlations, where 
the orbitals are $\exp(i \bm{k}\cdot \tilde{\bm{r}_i})$, with
\begin{equation}
\tilde{\bm{r}_i} = \bm{r}_i + \lambda \sum_{j \ne i} \eta(r_{ij}) \bm{r}_{ij} \ 
,
\label{backf1}
\end{equation}
$\eta(r)$ being a short-ranged function as a Gaussian~\citep{Manousakis83}. The 
calculation of excited states of $^3$He has also been possible by using CBF 
theory with increasing complexity due to Fermi statistics~\citep{Fabrocini88}. 
The high accuracy of CBF theory has been recently proved in a comparison 
between its predictions and very accurate experimental data on films of 
$^3$He~\citep{Godfrin12}. 

Working with the two isotopes of Helium, it is also possible to have stable 
mixtures of $^3$He and $^4$He, that is a Fermi-Bose mixture. In the limit of 
zero temperature, the mixture is stable up to a maximum $^3$He concentration of 
$~6.5$\% \citep{Ebner70}. This isotopic mixture was first studied neglecting the 
Fermi character of $^3$He atoms showing that the isotopic Bose-Bose mixture is 
not stable at any finite concentration~\citep{Tapash}. To get realistic results 
and finite 
solubilities it is mandatory to consider $^3$He atoms as fermions. The HNC/FHNC 
equations were generalized to the mixture by Fabrocini and Polls~\citep{Fabro82} 
and then extended to study the momentum distribution of the 
mixture~\citep{Boronat97}. The limiting case of a single impurity, known also as 
polaron, was also studied using HNC/CBF formalism both for the $^3$He impurity 
in bulk $^4$He~\citep{Boronat89} (Bose polaron) and for the $^4$He impurity in 
$^3$He~\citep{Arias94} (Fermi polaron).

\section{Quantum Monte Carlo methods}\label{qmc}

Variational methods discussed in the previous Section have been tremendously 
effective in the description of liquid Helium, with special significance on the 
accuracy achieved in the calculations of the dynamic response. However, the 
application of the variational principle guarantees upper bounds to the 
ground-state energies and it is difficult to know what remains between the 
bounds and exact values. Quantum Monte Carlo methods exploit the power of 
stochastic simulations to go beyond the bounds of the variational theories and 
obtain exact estimations within some statistical noise.

We first discuss the $T=0$ ground state with projection methods. The two 
methods that have been used in the study of liquid $^4$He are Green's function 
Monte Carlo (GFMC) and diffusion Monte Carlo (DMC). GFMC is the most accurate 
method because it has not any time-step dependence but its implementation is 
involved. Instead, DMC is a simpler method but with time-step dependence. 
Nowadays, DMC has become the most used tool. 

The DMC method solves the Schr\"{o}dinger equation, written 
in imaginary time,
\begin{equation}
- \frac{\partial \Psi(\bm{R},t)}{\partial t} = (H-E)\, \Psi(\bm{R},t) \ ,
\label{dmc.eq1}
\end{equation}
with $\bm{R} \equiv (\bm{r}_1,\ldots,\bm{r}_N)$ a $3N$-dimensional
vector ({\em walker}) and $t$ is the imaginary time. As it is usual in 
quantum mechanics, the time-dependent wave function of the system 
$\Psi(\bm{R},t)$ can be expanded in terms of a
complete set of eigenfunctions $\phi_i(\bm{R})$ of the Hamiltonian
\begin{equation}
\Psi(\bm{R},t)=\sum_{n}c_n \, \exp \left[\, -(E_i-E)t \, \right]\,
\phi_i(\bm{R})\ ,
\label{dmc.eq1b}
\end{equation}
$E_i$ being the eigenvalue associated to $\phi_i(\bm{R})$.  The
asymptotic solution of Eq.~(\ref{dmc.eq1}) for any value $E$ close to
the energy of the ground state, and for long times ($t \rightarrow
\infty$), gives $\phi_0(\bm{R})$, provided that there is a nonzero
overlap between $\Psi(\bm{R},t=0)$ and the ground-state wave function
$\phi_0(\bm{R})$.

A direct Monte Carlo implementation of Eq.~(\ref{dmc.eq1}) is hardly
able to work efficiently, especially in liquid Helium where the interatomic 
potential contains a hard core. To reduce the variance, one uses importance
sampling,  consisting in
rewriting the Schr\"{o}dinger equation in terms of the wave function
\begin{equation}
f(\bm{R},t)\equiv \psi(\bm{R})\,\Psi(\bm{R},t)\ ,
\label{dmc.eq2}
\end{equation}
where $\psi(\bm{R})$ is a time-independent trial wave function that describes
approximately the ground state of the system at the variational level.
Then, Eq.~(\ref{dmc.eq1}) turns out to be
\begin{equation}
-\frac{\partial f(\bm{R},t)}{\partial t}  =  -D\, 
 \bm{\nabla}^2_{\bm{R}}
f(\bm{R},t)+D\, \bm{\nabla}_{\bm{R}} \left( \bm{F}(\bm{R})
\,f(\bm{R},t)\,
\right)+\left(E_L(\bm{R})-E \right)\,f(\bm{R},t)  \ , 
\label{dmc.eq4}
\end{equation}
with $D=\hbar^2 /(2m)$, $E_L(\bm{R})=\psi(\bm{R})^{-1} H \psi(\bm{R})$
is the local energy, and
\begin{equation}
\bm{F}(\bm{R}) = 2\, \psi(\bm{R})^{-1} 
\bm{\nabla}_{\bm{R}} \psi(\bm{R})
\label{dmc.eq5}
\end{equation}
is called drift or quantum force. $\bm{F}(\bm{R})$ acts as an external force
which guides the diffusion process, involved by the first term in 
Eq.~(\ref{dmc.eq4}), to regions where $\psi(\bm{R})$ is large.

The r.h.s. of Eq.~(\ref{dmc.eq4}) may be written as the action of three
operators $A_i$ acting on the wave function $f({\bf R},t)$
\begin{equation}
-\frac{\partial f(\bm{R},t)}{\partial t} = (A_1+A_2+A_3)\, 
f(\bm{R},t) \equiv A\, f(\bm{R},t) \ .
\label{dmc.eq4p}
\end{equation}
The Schr\"odinger equation~(\ref{dmc.eq4p}) is written in a integral form 
by introducing the Green
function $G(\bm{R}^{\prime},\bm{R},t)$, which gives the
transition probability  from an initial state $\bm{R}$ to a final one 
$\bm{R}^{\prime}$ during a time $t$ 
\begin{equation}
     f(\bm{R}^{\prime},t+\Delta t) =\int G(\bm{R}^{\prime},\bm{R},
\Delta t)\, f(\bm{R},t)\, d \bm{R} \ .
\label{dmc.eq6}
\end{equation}
The DMC method
relies on reasonable approximations of $G(\bm{R}^{\prime},\bm{
R},\Delta t)$ for small values of the time-step $\Delta
t$. Considering such a short-time approximation, Eq.~(\ref{dmc.eq6})
is then iterated repeatedly until to reach the asymptotic regime
$f(\bm{R},t \rightarrow \infty)$, a limit in which one is effectively
sampling the ground state.

\begin{figure}
\begin{center}
\includegraphics[width=0.4\linewidth]{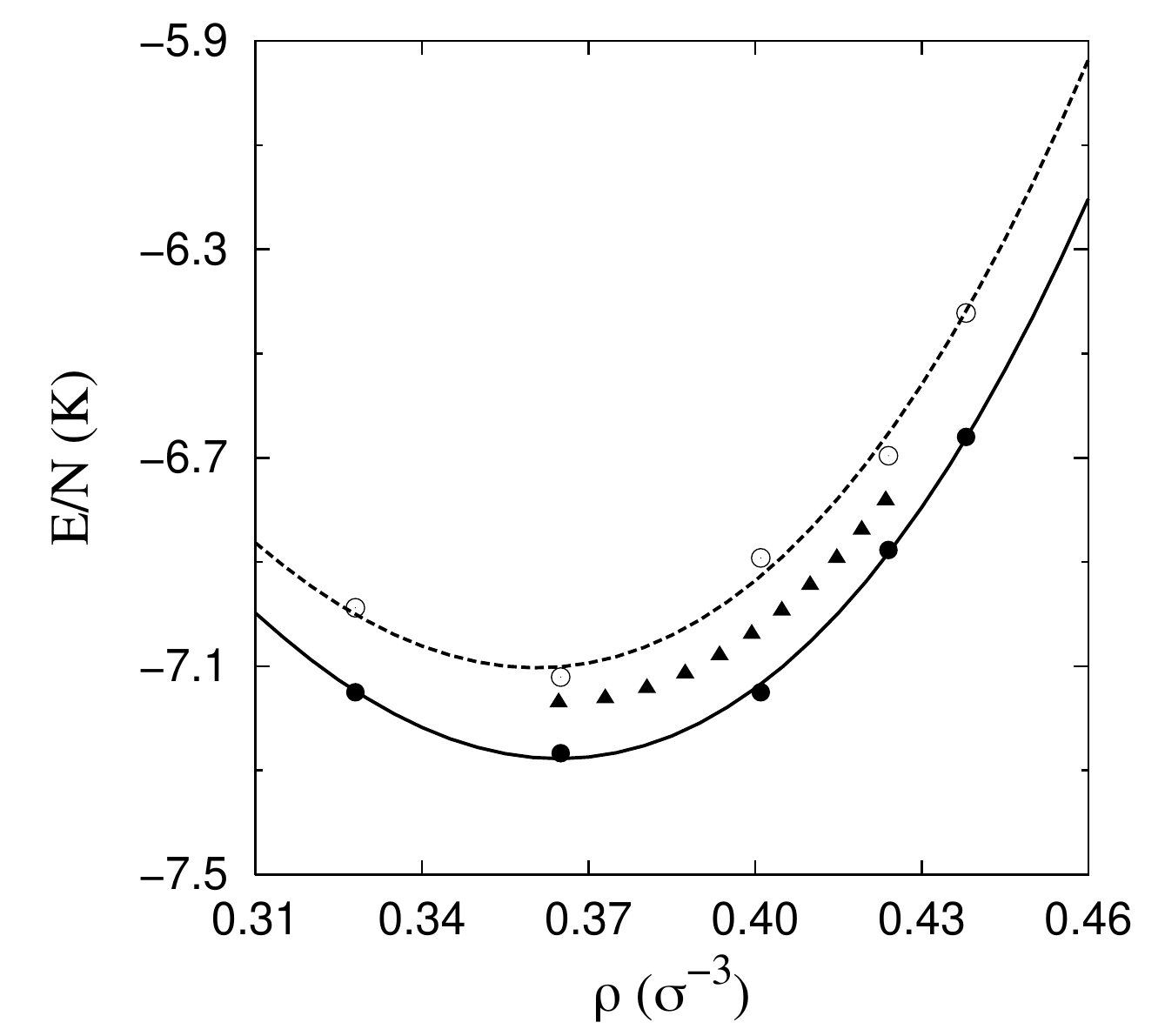}
\includegraphics[width=0.4\linewidth]{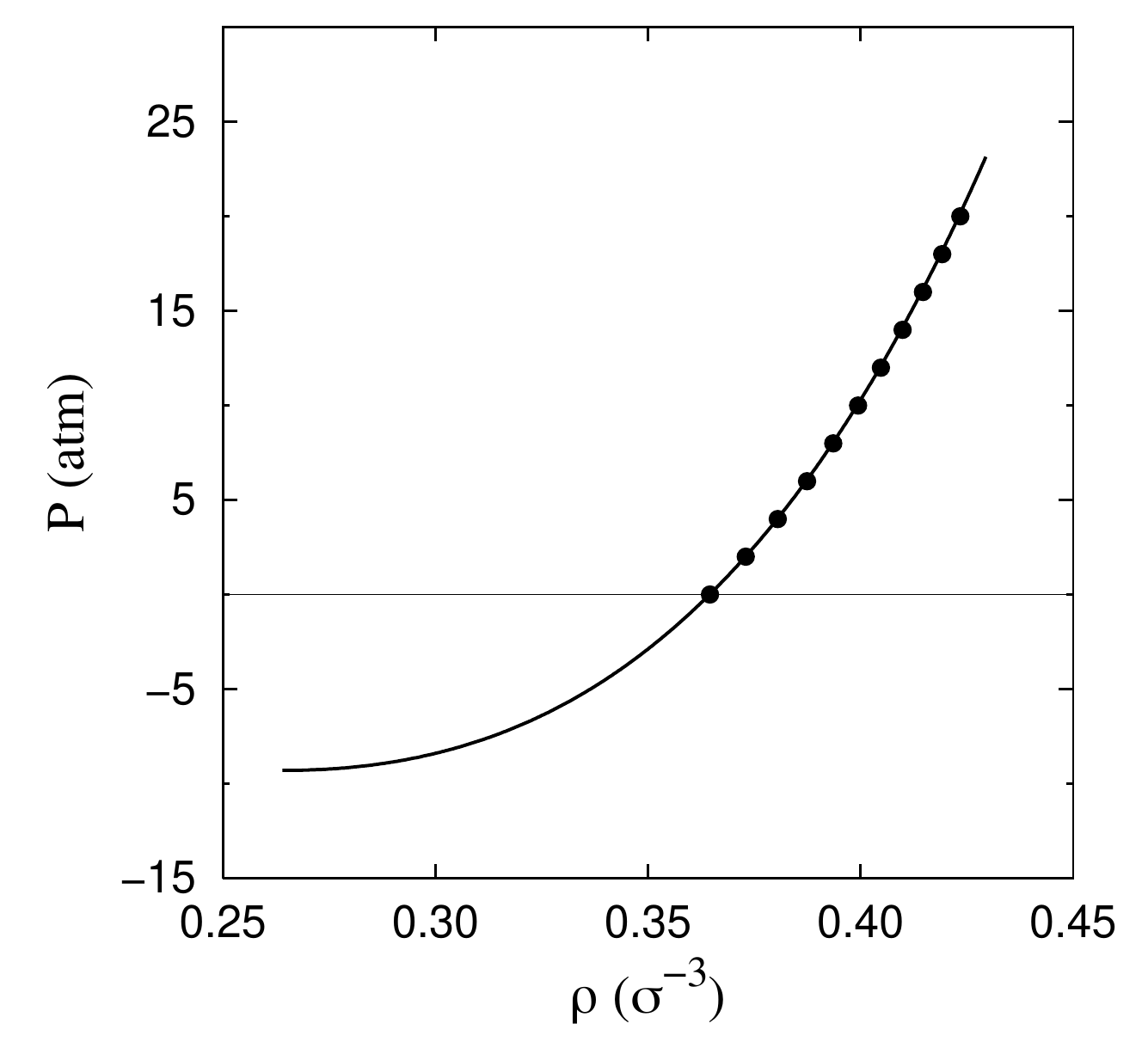}
\end{center}
\caption{\textit{Left:} Equation of state  of liquid $^4$He~\citep{Boro94}: 
open and full circles are DMC results with the Aziz~\citep{Aziz} and Aziz 
II~\citep{AzizII} potentials, respectively;  experimental values, solid 
triangles, from Ref.~\citep{bruyn}. 
\textit{Right:} Pressure as a function of density of liquid $^4$He:
points, experimental results~\citep{bruyn}, and solid line, DMC results. 
}
\label{fig-enerqmc}
\end{figure}

As it is shown in Fig. \ref{fig-enerqmc}, the accuracy achieved by the DMC 
method to reproduce the experimental equation of state is 
impressive~\citep{Boro94}. The dependence of the energy with the density is 
well reproduced, in spite of a nearly constant shift that it is related to the 
interatomic potential used in the simulations. We observe that the shift is 
practically constant looking at the dependence of the pressure with the density, 
as shown in Fig. \ref{fig-enerqmc}.

\begin{figure}
\begin{center}
\includegraphics[width=0.5\linewidth]{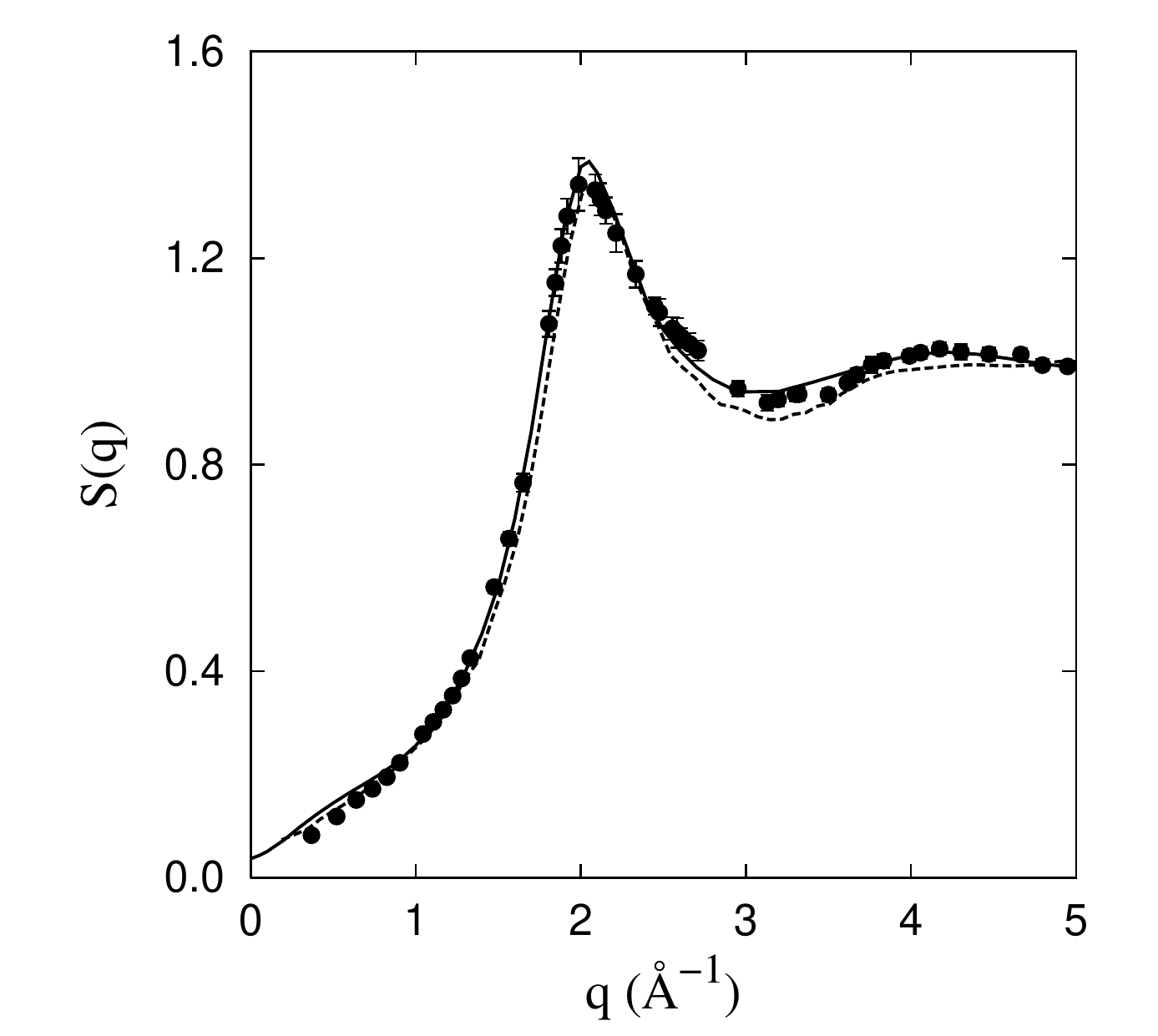}
\end{center}
\caption{DMC static structure function (points)
 for liquid $^4$He at $\rho_0^{\rm expt}$, in comparison with
experimental determinations from Refs.~\citep{svensson} (solid line)
 and \citep{wirth} (dashed
line). The error bars of the theoretical points are only shown where
larger than the size of the symbols.  }
\label{sqdmc}
\end{figure}

Quantum Monte Carlo methods are able to estimate other properties than the 
energy. However, direct estimations of operators which do not commute with the 
Hamiltonian are biased by the trial wave function used for importance sampling. 
To eliminate this bias it is necessary to work with pure 
estimators~\citep{casu_pures}, based on the forward-walking 
method~\citep{forward_walking}. In Fig. \ref{sqdmc}, we show DMC results for the 
static structure factor $S(q)$, calculated with pure estimators, in comparison 
with experimental data obtained from X-rays~\citep{wirth} and neutron 
scattering~\citep{svensson}. As one can see, the agreement between theory and 
experiment is excellent.

The study of the Fermi isotope $^3$He using Monte Carlo is much more involved 
due to the famous sign problem. In fact,
the Monte Carlo interpretation of the imaginary-time Schr\"odinger
equation requires that $f({\bf R},t)=\psi({\bf R}) \Psi({\bf R},t)$ be
a density, i.e., $\psi \, \Psi \geq 0$ in all the domain. This
boundary condition can be satisfied if $\psi$ and $\Psi$ change sign
together and thus share the same nodes. This approximation, known as
fixed-node (FN) method,~\citep{reynolds1} has been
extensively used in the ground-state calculations of
Fermi quantum liquids.
It can be proved that, due to that nodal constraint,
the fixed-node energies are variational upper bounds to the exact
eigenvalues for a given symmetry~\citep{reynolds1}.  Therefore, the FN results 
depend significantly on the {\em quality} of the trial wave function. The best 
upper bounds have been achieved by introducing backflow 
correlations, which move the nodal surface according to the interparticle 
interactions. Recently, an iterative backflow scheme has shown accurate results 
for the $^3$He equation of state~\citep{Holzmann}, in spite of being not so good 
as the ones obtained in $^4$He, where the calculation is not affected by the 
sign problem.

The thermal properties of liquid $^4$He have also been deeply studied using the 
path integral Monte Carlo (PIMC) method. As it is well known, 
the knowledge of the quantum partition function at finite temperature
\begin{equation}
Z = \text{Tr} \, e^{-\beta \hat{H}}
\label{zeta}
\end{equation}
allows for a full microscopic description of the properties of a given
system, with $\beta=1/T$ and $\hat{H} = \hat{K} + \hat{V}$ the Hamiltonian.
The non-commutativity of the quantum operators, kinetic $\hat{K}$ and 
potential $\hat{V}$ energies, makes
impractical a direct calculation of the partition function $Z$
(\ref{zeta}). Instead, all practical implementations intended for Monte
Carlo estimations of $Z$ rely on approximations that use as a starting
point the convolution property
\begin{equation}
e^{-\beta (\hat{K} + \hat{V})} = \left( e^{-\varepsilon ( \hat{K} + \hat{V}
)} \right)^M \ ,
\label{convolution}
\end{equation}
with $\varepsilon = \beta/M$, where  each one of the terms in the r.h.s.
corresponds to a higher temperature $M T$. In the most simple
approximation, known as primitive action (PA), the kinetic and potential
contributions factorize
\begin{equation}
e^{-\varepsilon ( \hat{K} + \hat{V})} \simeq e^{-\varepsilon  \hat{K}} \,
e^{-\varepsilon  \hat{V}} \ ,
\label{primitive}
\end{equation}
the convergence to the exact result being warranted by the Trotter
formula~\citep{trotter}
\begin{equation}
e^{-\beta (\hat{K} + \hat{V})}  = \lim_{M \to \infty} \left(
e^{-\varepsilon  \hat{K}}  \, e^{-\varepsilon  \hat{V}}  \right)^M \ .
\label{trotter}
\end{equation}

\begin{figure}
\begin{center}
\includegraphics[width=0.5\linewidth]{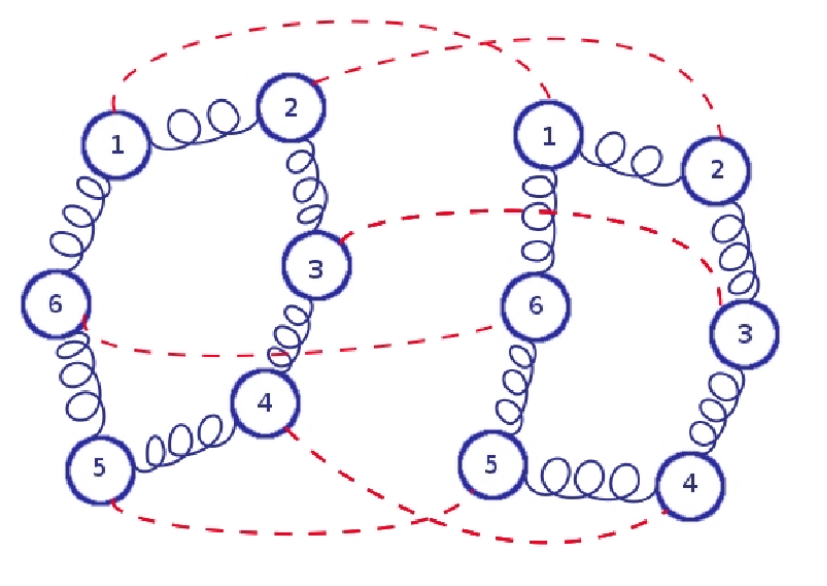}
\end{center}
\caption{Schematic representation of two particles in the PIMC formalism. The 
springs stand for the harmonic coupling between beads of every single atom 
(kinetic action) and the dashed lines for the interatomic potential between 
beads of different atoms. }
\label{springs}
\end{figure}

The primitive action is a particular case of a more general expansion in which 
one can decompose the action, that is
\begin{equation}
e^{-\varepsilon (\hat{K} + \hat{V})} \simeq \prod_{i=1}^{n} e^{-t_i
\varepsilon \hat{K}} \, e^{-v_i \varepsilon \hat{V}} \ ,
\label{suzukin}
\end{equation}
with $\{ t_i,v_i \}$ parameters to be determined. A particular implementation 
of this expansion put forward by Chin~\citep{chin} has proved to be a full 
fourth-order approximation that even can behave as sixth order by tuning 
properly its free parameters~\citep{pimc_chin}. Alternatively, one can use the 
pair-action approximation, in which the total action is built as product of 
pair actions that are obtained  either exactly or with some 
approximations~\citep{Ceperley_rmp}. 
Considering distinguishable particles, the 
quantum partition
function $Z$ (\ref{zeta}) can be obtained through a multidimensional
integral of the $M$ terms (beads) in which it is decomposed,
\begin{equation}
Z = \int d \bm{R}_1 \ldots d \bm{R}_M \ \prod_{\alpha=1}^{M} \rho
(\bm{R}_\alpha,\bm{R}_{\alpha+1}) \ ,
\label{zeta2}
\end{equation}
with $\bm{R} \equiv \{\bm{r}_1,\ldots,\bm{r}_N\}$ and $\bm{R}_{M+1}
=\bm{R}_1$. The PIMC method is then mapped to classical closed polymers 
composed by beads harmonically coupled by the kinetic action (Fig. 
\ref{springs}).

To ensure the  Bose symmetry of $^4$He it is necessary to symmetrize the 
quantum partition function (\ref{zeta}). This requires of an additional 
sampling in the permutation space. The first 
applications of PIMC to liquid $^4$He were made by proposing permutations 
between pairs, triplets, and more particles~\citep{Ceperley_rmp}. However, this 
method becomes inefficient when the total number of particles in the simulation 
increases. A significant better performance has been achieved with the 
introduction of the worm algorithm, first proposed in lattice 
systems~\citep{Prokofev}, and then extended to continuum 
fluids~\citep{Boninsegni}.

\begin{figure}
\begin{center}
\includegraphics[width=0.5\linewidth]{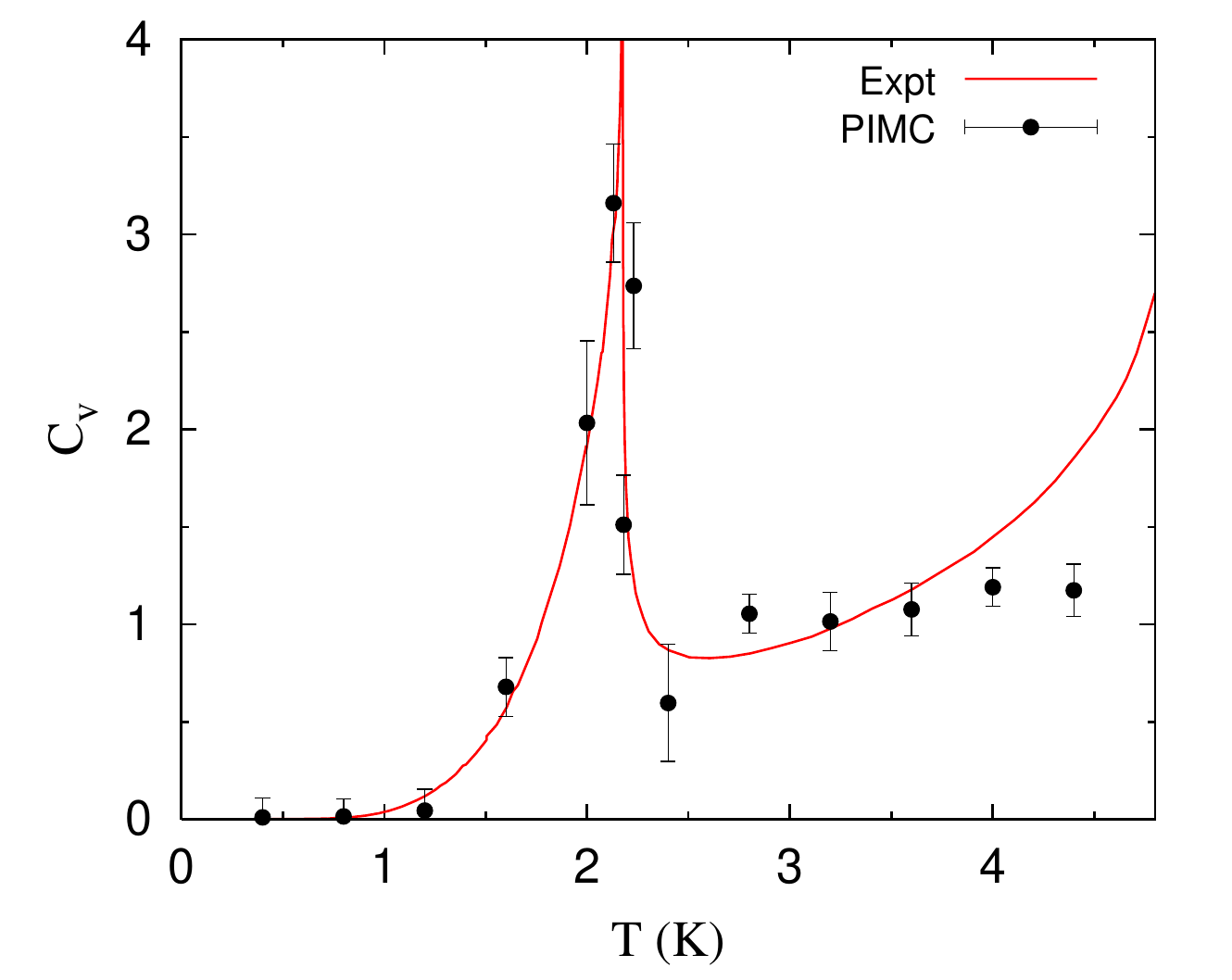}
\end{center}
\caption{Specific heat of liquid $^4$He obtained with the PIMC method around 
the normal-to-superfluid $\lambda$ phase transition. The line corresponds to 
experimental results~\citep{Wilks67}.}
\label{heat}
\end{figure}

One of the more impressive successes of PIMC was the characterization of the 
normal-to-superfluid $\lambda$ phase transition at a critical temperature 
$T_\lambda=2.17$ K. In Fig. \ref{heat}, we compare PIMC results of the specific 
heat as a function of $T$  with experimental data. The estimation 
of the specific heat is numerically involved but its singularity at $T_\lambda$ 
is well reproduced.

\section{Summary}
We have reported in this article some of the main theoretical approaches used 
in the study of liquid Helium. After an Introduction on the physical 
properties of liquid Helium, we have discussed the early theoretical 
approaches used in its study. In the third Section, we have discussed on 
the application of the variational theory and correlated basis function as 
a best approach for its microscopic understanding. In the fourth Section, we 
have introduced the main quantum Monte Carlo methods used in the study of 
liquid Helium. 

This subject has been running for many years and 
it is impossible in practice to include all the immense work developed. Liquid 
Helium has been a continuous benchmark for quantum many-body theories and has 
been the workhorse for improving theories at all levels. This continued effort 
has produced a set of very accurate theoretical tools which have dramatically 
reproduced many experimental data. Liquid Helium is a really very strongly 
correlated quantum fluid where any simple perturbative scheme cannot be used. 
Remarkably, variational theory has proved to be the most efficient tool to deal 
with its microscopic description. Finally, it is also noticeable how 
progress in the development of quantum Monte Carlo methods has grown stimulated 
for the incredible amount of experimental data that we have at hands.   

\section*{Acknowledgments}
We acknowledge financial support from Ministerio de
Ciencia e Innovaci\'on  MCIN/AEI/10.13039/501100011033
(Spain) under Grant PID2020-113565GB-C21.











\bibliographystyle{apsr} 

\bibliography{encyclop}


\end{document}